\documentclass[12pt]{article}
\usepackage[cp1251]{inputenc}

\usepackage{amsfonts,amssymb,amsthm,mathtools}
\usepackage{amsmath}
\usepackage{icomma}
\usepackage{latexsym}
\usepackage{graphicx}
\usepackage{subfigure}
\usepackage {bm}

\usepackage{geometry} 
\begin{document}
	
	\begin{center}
		\textbf{THE POSSIBILITY TO EXPERIMENTALLY DETERMINE THE STRUCTURE OF A FERMIONIC VACUUM IN QUANTUM ELECTRODYNAMICS}
	\end{center}
	
		\begin{center}
		
		{V.~P.~Neznamov\footnote{vpneznamov@mail.ru, vpneznamov@vniief.ru}}\\
		
		\hfil
		{\it \mbox{	Russian Federal Nuclear Center--All-Russian Research Institute of Experimental Physics},  Mira pr., 37, Sarov, 607188, Russia} \\
	\end{center}

\begin{abstract}
	\noindent
	\footnotesize{In the standard quantum electrodynamics (QED), the fermionic vacuum is a 
		continuum of randomly created and annihilated virtual electron-positron 
		pairs. In this case, in the strong electromagnetic fields,vacuum creation of 
		real electron-positron pairs is possible. In particular, in the standard QED 
		in a strong uniform electrical field, the Schwinger effect is implemented.
		Currently, there exist the QED versions with empty fermionic vacuum without 
		fluctuations of creation and annihilation of virtual electron-positron 
		pairs. These versions are the (QED)$_{FW}$ in the Foldy-Wouthuysen 
		representation, the (QED)$_{KG}$ with spinor equations of the Klein-Gordon 
		type, the (QED)$_{DN}$ with opposite signs in front of particle and 
		antiparticle masses in Dirac equations and with the use of only states with 
		positive energies in S-matrix elements. The latter relates to both real and 
		virtual energy states.
		In the paper, we propose to carry out a set of experiments at colliders with 
		collisions of heavy ions to determinethe nature of the fermionic vacuum. The 
		measurements of the emission of electron-positron pairs depending on the 
		total charge of colliding ions $\left( {Z_{\Sigma } =146\div 184} \right)$ 
		show the structure of a fermionic vacuum in quantum electrodynamics.} \\
	
	
	
\end{abstract}

\section{Introduction}	

The state-of-the-art concept of the fermionic vacuum can be expressed by the 
following quotation: ''{\it Relativistic quantum mechanics and quantum field theory with the possibility of pair creation laid the grounds for our present concept of the nature of a vacuum. In today's language, a vacuum consists of a polarizable gas of virtual particles, fluctuating randomly. It is found that in the presence of strong external fields, the vacuum may even contain "real" particles}''\cite{bib1}.

In the standard QED, in strong electromagnetic fields, vacuum creation of 
real electron-positron pairs is possible. In particular, the Schwinger 
effect is implemented in a strong uniform electrical field \cite{bib2}.

Currently, there are QED versions with the empty fermionic vacuum without the polarizable ''gas'' of virtual particles:

\begin{itemize}
	\item the (QED)$_{FW}$ in the Foldy-Wouthuysen representation \cite{bib3} - \cite{bib5},
	\item the (QED)$_{KG}$ with spinor equations of the Klein-Gordon type \cite{bib6, bib7},
	\item the (QED)$_{DN}$ with opposite signs in front of particle and antiparticle masses in Dirac equations and with the use of only real and virtual states with positive energies in S-matrix elements \cite{bib8}.
\end{itemize}
In QED versions with an empty fermionic vacuum, there is no vacuum creation 
of electron--positron pairs.Within the framework of the perturbation theory, 
the observed effects of quantum electrodynamics are described in the same 
way as in the standard QED. In this case, it is sufficient to use solutions 
with positive energies of fermions in calculations. It refers to both real 
and virtual intermediate states.

Can we determine the fermionic vacuumstructure experimentally? We answer this 
question positively, proposing a set of experiments with collisions of heavy 
ions with the total $Z_{\Sigma } =Z_{1} +Z_{2} =146\div 184$ at the 
colliders of NICA \cite{bib9}, FAIR \cite{bib10}, HIAF \cite{bib11}. The 
aim of the experiments is the measurement of emitted electron-positron pairs 
$P\left( {Z_{\Sigma } } \right)$ as a function of $Z_{\Sigma } $. For the 
standard QED with the nonempty fermionic vacuum, the dependence $P_{st} \left( {Z_{\Sigma } } \right)$ is a smooth continuous function. For the QED 
versions with empty fermionic vacuum, the numbers of $P_{FW} \left( 
{Z_{\Sigma } } \right)$ become lower; the relationship $P_{FW} \left( 
{Z_{\Sigma } } \right)$ is a step function. The reasons for such a behavior 
of therelationships $P\left( {Z_{\Sigma } } \right)$ are the differences in 
the contributions of electron negative-energy levels.

For the standard QED in the range of $Z_{\Sigma } =147\div 170$, there exists 
the negative-energy level $1S_{1/2} $, and in the range of $Z_{\Sigma } =169\div 183$, there exists the negative-energy level $2P_{1/2}$ \cite{bib1}. For the QED versions with an empty fermionic 
vacuum, in finite expressions, we deal only with positive energies and there 
is no physical contribution of the negative-energy levels $1S_{1/2}$ 
and $2P_{1/2}$. It leads to the essential decrease in the 
number of emitted electron-positron pairs within the range of $Z_{\Sigma } 
=146\div 184$ and to the step nature of the relationship $P_{FW} \left( 
{Z_{\Sigma } } \right)$.

So, after carrying-out the proposed set of experiments, we will arrive at 
the unambiguous understanding of the fermionic vacuum structure in QED. 

\section{The nonperturbative standard QED in the fields of hydrogen-like ions with a large charge number $Z$}
For the standard QED with the fluctuating fermionic vacuum, the low-lying 
energy levels of hydrogen-like ions are presented as the function of the 
nuclear charge number $Z$ (see Fig. 1). The figure is from Ref. \cite{bib12}.


\begin{figure}[h!]
	\centerline{\includegraphics[width=0.6\linewidth]{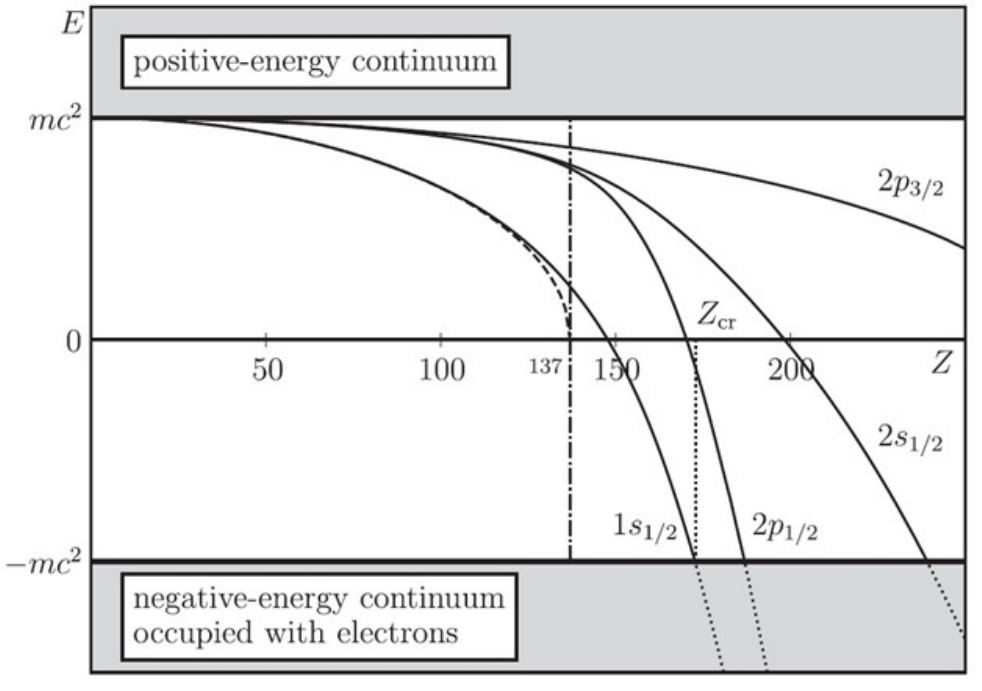}}
	\caption{The low-lying energy levels of the hydrogen-like ion as the function 
		of the nuclear charge number $Z$.}
	\label{ris:Fig.1}
\end{figure}


Let us consider the level $1S_{1/2}$. For the Coulomb field of the nucleus point 
charge $+Ze$, the electron level $1S_{1/2}$ disappears at $Z=137$. If we take 
into account the finite dimensions of atomic nuclei \cite{bib13} - \cite{bib16}, the energy of 
state $1S_{1/2}$ becomes negative at $Z>146$. At $Z_{cr} 
\approx 171$, the level $1S_{1/2}$ passes to the negative-energy continuum. 
Similarly, the energy of level $2P_{1/2}$ becomes negative at $Z>168$; at $Z_{cr} \approx 184$, the level $2P_{1/2}$ passes to the negative- energy continuum. 
According to the theoretical predictions of the standard QED, when a level 
passes to the negative-energy continuum, a neutral vacuum disintegrates 
emitting two electron-positron pairs \cite{bib14, bib15}.

The stable and quasi-stable nuclei with $Z>118$ do not exist in nature yet. 
In order to study QED effects, there are proposals to use 
quasi-molecules generated over a some time period incollisions of heavy ions with the total $Z_{\Sigma } =Z_{1} +Z_{2} $ \cite{bib17}-\cite{bib33}. In Fig. 2, the low-lying energy levels as a 
function of time are presented for a quasi-molecule generated in the 
collision of two uranium ions. The creation of dynamic (a, b, d, e, f 
arrows) and vacuum (c arrow) electron-positron pairs is also schematically 
shown ibidem. The dynamic creation of pairs occurs at any $Z$. The creation of 
pairs due to disintegration of a vacuum is especially prominent at $Z\ge
Z_{cr}$ (for the level $1S_{1/2}$, $Z_{cr} \approx 171$; for the level $2P_{1/2}$, $Z_{cr} \approx 184$ etc.). This figure is from Ref. \cite{bib12}.


\begin{figure}[h!]
	\centerline{\includegraphics[width=0.6\linewidth]{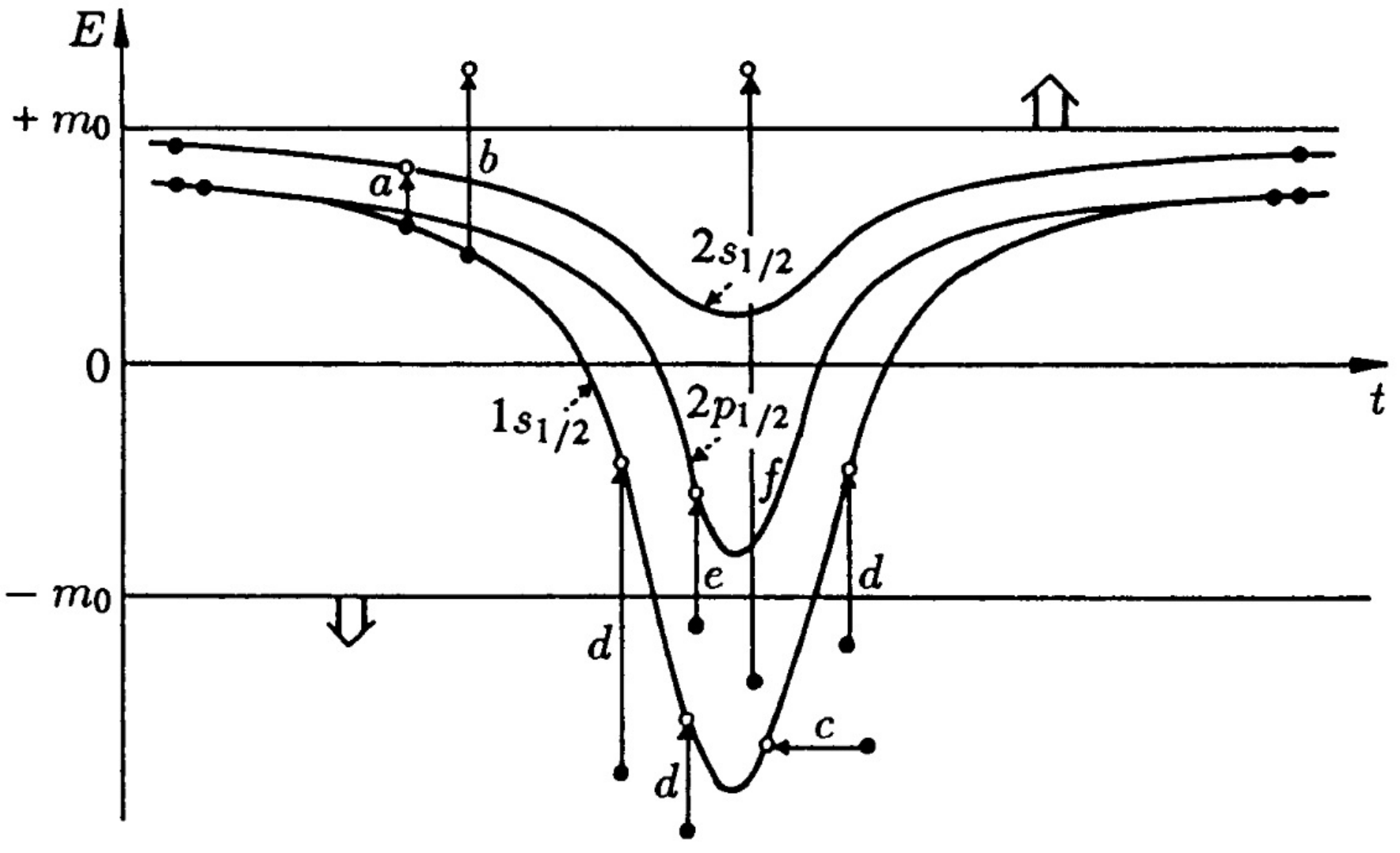}}
	\caption{The low-lying energy levels of a quasi-molecule as a function of time.}
	\label{ris:Fig.2}
\end{figure}

\section{Nonperturbative QED versions with empty fermionic vacuumin strong electrical fields}
Within the framework of the perturbation theory in the QED versions with the empty fermionic vacuum \cite{bib3}, \cite{bib5}-\cite{bib8}), when physical 
effects are calculated, the transition amplitudes with involvement of 
negative-energy states are not present. In the transformed Dirac equations, 
we use the states with positive energies of the Dirac equation for positrons 
instead of states with electron negative energies. The states with negative 
energies are taken into account just to fulfill mathematical conditions of 
completeness in expansions of quantum-mechanical operators and wave 
functions.

Below, we will consider the validation of the above description for the 
versions of the nonperturbative QED with empty fermionic vacuum.


\subsection{Equations of the nonperturbative QED in Feynman-Gell-Mann and Foldy-Wouthuysen representations}

In the recent paper \cite{bib34}, in the FW and (FG) 
representations, we obtained equations with closed expressions of the energy 
operator in the presence of the Coulomb field $eA^{0}\left( r \right)$. A brief overview of the FW and FG representations is provided in the Appendix. 
Below, we will consider the Coulomb potential of ionized nuclei $\left( 
{A^{0}\left( r \right)\sim \left| e \right|Z} \right)$.

In compliance with Ref. \cite{bib34}, let us write some of the 
equations for electrons and positrons.\footnote{Below, the system of units 
	$\hslash =c=1$ is used; $p^{\mu }=i\frac{\partial }{\partial x_{\mu } 
	};\,\,\,\mu =0,1,2,3;\,\,\,{\rm {\bf \pi }}({\rm {\bf x}},t)={\rm {\bf 
			p}}-e{\rm {\bf A}}({\rm {\bf x}},t)$, $A^{\mu }({\rm {\bf x}},t)$ are 
	electromagnetic potentials; $\alpha^{i},\,\,\beta =\gamma^{0}$ are 
	four-dimensional Dirac matrices are in the standard representation $\left( 
	{i=1,2,3} \right)$, and $\gamma^{i} =\gamma^{0} \alpha^{i} $, 
	$\Sigma^{i} =\left( {\begin{array}{l}
			\sigma^{i}\;\;\,0{\kern 1pt} \\ 
			0\;\;\;\sigma^{i} \\ 
	\end{array}} \right)$, $\sigma^{i}$ are two-dimensional Pauli matrices 
	$\left( {i=1,2,3} \right)$.}
\begin{enumerate}
	\item The equation for electrons with positive energies:
	\[
	\left( {\varepsilon =\left| E \right|>0,\,\,\,e=-\left| e \right|<0} 
	\right),
	\]
	\begin{equation}
		\label{eq1}
		\left( {\left( {\left| E \right|+\left| e \right|A^{0}} \right)^{2}-{\rm 
				{\bf p}}^{2}-m^{2}-i\left| e \right|{\rm {\bm \sigma }}\,\nabla A^{0}} 
		\right)\varphi_{FG}^{e} =0.
	\end{equation}
	\item The equation for electrons with negative energies
	\[
	\left( {\varepsilon =-\left| E \right|<0,\,\,\,e=-\left| e \right|<0} 
	\right),
	\]
	\begin{equation}
		\label{eq2}
		\left( {\left( {\left| E \right|-\left| e \right|A^{0}} \right)^{2}-{\rm 
				{\bf p}}^{2}-m^{2}+i\left| e \right|{\rm {\bm \sigma }}\,\nabla A^{0}} 
		\right)\chi_{FG}^{e} =0.
	\end{equation}
	\item The equation for positrons with positive energies
	\[
	\left( {\varepsilon =\left| E \right|>0,\,\,\,e=\left| e \right|>0} \right),
	\]
	\begin{equation}
		\label{eq3}
		\left( {\left( {\left| E \right|-\left| e \right|A^{0}} \right)^{2}-{\rm 
				{\bf p}}^{2}-m^{2}+i\left| e \right|{\rm {\bm \sigma }}\,\nabla A^{0}} 
		\right)\varphi_{FG}^{p} =0.
	\end{equation}
	\end{enumerate}

Note that Eqs. (\ref{eq1}) - (\ref{eq3}) were derived in Ref. \cite{bib34} using FW unitary transformations. The solutions to Eqs. 
(\ref{eq1}) - (\ref{eq3}) are also solutions of the corresponding original Dirac equations.

For electrons with energy $\varepsilon >0$, $\left( {\psi_{FW}^{\left( + 
		\right)} \left( {{\rm {\bf x}},t} \right)} \right)^{e}=\left( 
{U_{FW}^{\left( + \right)} \left( {{\rm {\bf x}}} \right)} \right)^{e}\left( 
{\psi_{D}^{\left( + \right)} \left( {{\rm {\bf x}},t} \right)} 
\right)^{e}$;

for electrons with energy $\varepsilon <0$, $\left( {\psi_{FW}^{\left( - 
		\right)} \left( {{\rm {\bf x}},t} \right)} \right)^{e}=\left( 
{U_{FW}^{\left( - \right)} \left( {{\rm {\bf x}}} \right)} \right)^{e}\left( 
{\psi_{D}^{\left( - \right)} \left( {{\rm {\bf x}},t} \right)} 
\right)^{e}$;

for positrons with energy $\varepsilon >0$, $\left( {\psi_{FW}^{\left( + 
		\right)} \left( {{\rm {\bf x}},t} \right)} \right)^{p}=\left( 
{U_{FW}^{\left( + \right)} \left( {{\rm {\bf x}}} \right)} \right)^{p}\left( 
{\psi_{D}^{\left( + \right)} \left( {{\rm {\bf x}},t} \right)} 
\right)^{p}$.

An important result established in Ref. \cite{bib34} is the 
connection between the FW and FG representations (see expressions (\ref{eq4}) - (\ref{eq6})). This connection 
automatically resolves the problem of ''extraneous'' solutions in the 
FG equations (see Appendix A).

In Eq. (\ref{eq1}), the spinor $\varphi_{FG}^{e} \left( {{\rm {\bf x}}} 
\right)$ in the FG representation is proportional to the upper 
spinor $\varphi_{c}^{e} \left( {{\rm {\bf x}}} \right)$ in the 
FW representation
\begin{equation}
	\label{eq4}
	\left( {\varphi_{FG}^{e} \left( {{\rm {\bf x}}} \right)=A_{\left( + 
			\right)}^{e} \varphi_{c}^{e} \left( {{\rm {\bf x}}} \right);\,\,\,\left( 
		{\psi_{FW}^{\left( + \right)} \left( {{\rm {\bf x}}} \right)} 
		\right)^{e}=\left( {{\begin{array}{*{20}c}
					{\varphi_{c}^{e} \left( {{\rm {\bf x}}} \right)} \hfill \\
					\,\,\,\,\,\,0 \hfill \\
		\end{array} }} \right)} \right).
\end{equation}
In Eq. (\ref{eq2}), the spinor $\chi_{FG}^{e} \left( {{\rm {\bf x}}} \right)$ in 
the FG representation is proportional to the lower spinor 
$\chi_{c}^{e} \left( {{\rm {\bf x}}} \right)$ in the FW 
representation
\begin{equation}
	\label{eq5}
	\left( {\chi_{FG}^{e} \left( {{\rm {\bf x}}} \right)=A_{\left( - 
			\right)}^{e} \chi_{c}^{e} \left( {{\rm {\bf x}}} \right);\,\,\,\left( {\psi 
			_{FW}^{\left( - \right)} \left( {{\rm {\bf x}}} \right)} \right)^{e}=\left( 
		{{\begin{array}{*{20}c}
				\,\,\,\,\,\,\,0 \hfill \\
					{\chi_{c}^{e} \left( {{\rm {\bf x}}} \right)} \hfill \\
		\end{array} }} \right)} \right).
\end{equation}
In Eq. (\ref{eq3}), the spinor $\varphi_{FG}^{p} \left( {{\rm {\bf x}}} 
\right)$ in the FG representation is proportional to the upper 
spinor $\varphi_{c}^{p} \left( {{\rm {\bf x}}} \right)$ in the 
FW representation
\begin{equation}
	\label{eq6}
	\left( {\varphi_{FG}^{p} \left( {{\rm {\bf x}}} \right)=A_{\left( + 
			\right)}^{p} \varphi_{c}^{p} \left( {{\rm {\bf x}}} \right);\,\,\,\left( 
		{\psi_{FW}^{\left( + \right)} \left( {{\rm {\bf x}}} \right)} 
		\right)^{p}=\left( {{\begin{array}{*{20}c}
					{\varphi_{c}^{p} \left( {{\rm {\bf x}}} \right)} \hfill \\
					\,\,\,\,\,\,0 \hfill \\
		\end{array} }} \right)} \right).
\end{equation}
Above,

$$A_{\left( + \right)}^{e} =\left( {1+\dfrac{m^{2}}{\left( {\left| E 
			\right|+{\rm {\bm \sigma \bf {p}}}+\left| e \right|A^{0}} \right)^{2}}} 
\right)^{-1/ 2},\,\,\, \\ $$
$$A_{\left( - \right)}^{e} =\left( 
{1+\dfrac{m^{2}}{\left( {\left| E \right|+{\rm {\bm \sigma \bf {p}}}-\left| e 
			\right|A^{0}} \right)^{2}}} \right)^{-1/ 2},\,\,\, \\ $$
		$$A_{\left( + \right)}^{p} =\left( 
{1+\dfrac{m^{2}}{\left( {\left| E \right|+{\rm {\bm \sigma \bf {p}}}-\left| e 
			\right|A^{0}} \right)^{2}}} \right)^{-1/2}.$$

As a result, we see that Eq. (\ref{eq3}) for positrons with positive 
energies, $\varepsilon >0$, coincides with Eq. (\ref{eq2}) for electrons with 
negative energies, $\varepsilon <0$.

In our case, the positrons are in the repulsive Coulomb field of ionized 
nuclei. For them, the upper continuum is available with the continuous 
energy spectrum $\varepsilon >m$. In this case, there are no stationary 
bound states with $\varepsilon <m$.

The equality of Eqs. (\ref{eq2}) and (\ref{eq3}) implies that the equivalent (accurate 
to the sign) continuous energy spectrum exists for electrons with negative 
energies and positrons with positive energies, with the identical stationary 
wave functions. However, in the spectrum of Eq. (\ref{eq2}) in the range of 
$Z_{\Sigma } =147\div 170$ , there is the negative energy level $1S_{1/2}$ and 
in the range of $Z_{\Sigma } =169\div 183,$ there is the negative energy 
level $2P_{1/2}$ (see Fig. 1).

For Eq. (\ref{eq3}), the existence of such bound states is impossible for simple 
physical reasons. Hence, the existence of bound states with negative energies 
is a mathematical artifact.

Since Eqs. (\ref{eq1}) - (\ref{eq3}) are obtained by unitary transformations of the 
Dirac equation, the conclusion about the absence of a physical (as opposed 
to mathematical) contribution from stationary bound states with negative 
energies to the calculated QED effects also holds for the original Dirac 
equation.

Let us note that Eq. (\ref{eq3}) with the changed sign in front of $A^{0}\left( 
{{\rm {\bf x}}} \right)$ (the motion of a positron in the attractive Coulomb 
field) coincides with Eq. (\ref{eq1}) for electrons. As it should be, discrete 
and continuous energy spectra of electrons and positrons, moving in the 
attractive Coulomb field, coincide with each other.

As in the domain of perturbation theory, in nonperturbative versions of QED 
with an empty fermionic vacuum, negative-energy states should not be 
included in calculations of physical effects. The positive-energy 
dependencies $E\left( Z \right)$ in this case are shown in Fig. 3. The 
difference between the spectra in Figs. 1 and 3 should lead to a 
noticeable change in the number of real electron-positron pairs produced.

Let us consider in more detail how probabilities of electron-positron pairs 
$P\left( Z \right)$ are calculated in the standard nonperturbative QED and 
how it is possible to calculate $P\left( Z \right)$ in the nonperturbative 
QED versions with empty fermionic vacuum, by using the developed formalism 
in the standard QED.


\begin{figure}[h!]
	\begin{minipage}[h]{0.47\linewidth}
		\center{\includegraphics[width=1\linewidth]{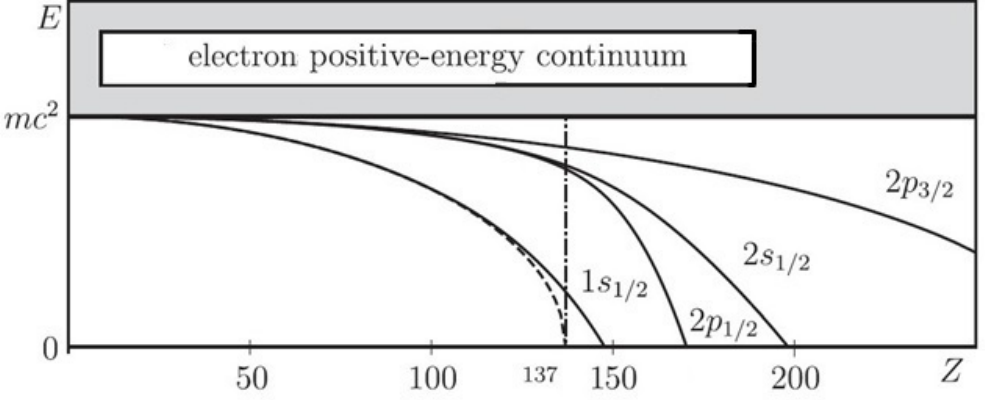}} a) \\
	\end{minipage}
	\hfill
	\begin{minipage}[h]{0.47\linewidth}
		\center{\includegraphics[width=1\linewidth]{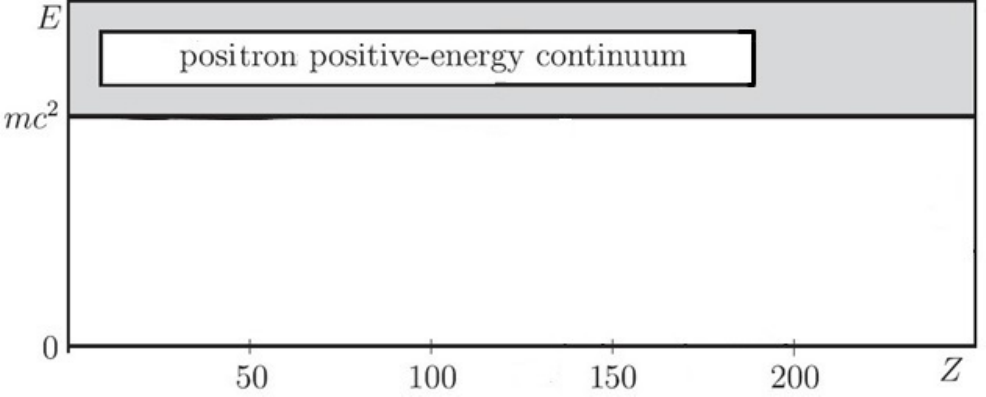}} \\b) 
	\end{minipage}
		\caption{The energy spectrum: a) equation for electrons (\ref{eq1}), b) equation for positrons (\ref{eq3}).}
	\label{ris:Fig.3}
\end{figure}


\section{Creation of electron-positron pairs in the nonperturbative standard QED}
\label{sec:creation}
Below, we present extracts from the formalism of papers by Shabayev's team (see Refs. \cite{bib35} - \cite{bib40}).

To determine the probabilities of pair creation, initially, we should 
consider the solutions of the Dirac equation in the external potential 
depending on time. The potential is induced by colliding nuclei whose motion 
is described by classical Rutherford trajectories. The Dirac equation has the form: 
\begin{equation}
	\label{eq7}
	i\frac{\partial }{\partial t}\psi \left( {{\rm {\bf r}},t} \right)=H\left( t 
	\right)\psi \left( {{\rm {\bf r}},t} \right),
\end{equation}
where
\begin{equation}
	\label{eq8}
	H\left( t \right)={\rm {\bm \alpha \bf {p}}}+\beta m+V\left( {{\rm {\bf r}},t} 
	\right).
\end{equation}
Here, ${\rm {\bm \alpha }}, \beta $ are Dirac matrices, $m$ is electron mass, 
$V\left( {{\rm {\bf r}},t} \right)$ is a two-center potential, induced by nuclei.

In (\ref{eq7}), (\ref{eq8}) and below, the system of units of $\hslash =c=1$ is used. In Refs. \cite{bib35} - \cite{bib40}, the 
monopole approximation is used for the potential $V\left( {{\rm {\bf r}},t} \right)$
\begin{equation}
	\label{eq9}
	V^{\left( {mon} \right)}\left( {r,t} \right)=\frac{1}{4\pi }\int {d\Omega 
		V\left( {{\rm {\bf r}},t} \right).} 
\end{equation}
With the spherically symmetric potential $V^{\left( {mon} \right)}\left( 
{r,t} \right)$, the Dirac equation (\ref{eq7}) allows separation of variables.

The Dirac wave functions are presented in the standard form
\begin{equation}
	\label{eq10}
	\psi_{\kappa m_{\varphi } } \left( {{\rm {\bf r}},t} 
	\right)=\frac{1}{r}\left( {{\begin{array}{*{20}c}
				\,\,{G_{\kappa } \left( {r,t} \right)\Omega_{\kappa m_{\varphi } } \left( 
					{\theta ,\varphi } \right)} \hfill \\
				{iF_{\kappa } \left( {r,t} \right)\Omega_{-\kappa m_{\varphi } } \left( 
					{\theta ,\varphi } \right)} \hfill \\
	\end{array} }} \right),
\end{equation}
where $\Omega_{\kappa m_{\varphi } } \left( {\theta ,\varphi } 
\right),  \Omega_{-\kappa m_{\varphi } } \left( {\theta ,\varphi } 
\right)$ are spherical spinors, $\kappa =\left( {-1} 
\right)^{j+l+\frac{1}{2}}\left( {j+\frac{1}{2}} \right)$ is the relativistic 
quantum number, determined by total and orbital moments $j,l$; $m_{\varphi } 
=-j,-j+1,...j$ is the projection of the total moment $j$.

As the result, we can write the radial Dirac equation for a specified 
value $\kappa $
\begin{equation}
	\label{eq11}
	i\frac{\partial }{\partial t}\Phi \left( {r,t} \right)=H_{\kappa } \left( t 
	\right)\Phi \left( {r,t} \right),
\end{equation}
where

\begin{equation}
	\Phi \left( {r,t} \right)=\left( {{\begin{array}{*{20}c}
				{G_{\kappa } \left( {r,t} \right)} \hfill \\
				{F_{\kappa } \left( {r,t} \right)} \hfill \\
	\end{array} }} \right).
\end{equation}

Here $\Phi \left( {r,t} \right)$ is the radial Dirac wave function.

$H_{\kappa } \left( t \right)=\left( {{\begin{array}{*{20}c}
			{m+V^{\left( {mon} \right)}\left( {r,t} \right)} \hfill & 
			\,\,\,\,\,\,\,\,\,\,{-\dfrac{d}{dr}+\dfrac{\kappa }{r}} \hfill \\
			\,\,\,\,\,\,\,\,\,\,\,{\dfrac{d}{dr}+\dfrac{\kappa }{r}} \hfill & {-m+V^{\left( {mon} 
					\right)}\left( {r,t} \right)} \hfill \\
\end{array} }} \right)$ is the matrix of the Dirac Hamiltonian in (\ref{eq11}).

Below, we will consider the quantum dynamics from the initial time $t_{in}$ up to the final time $t_{out}$.

Let us determine two set of solutions for Dirac equation (\ref{eq7}) $\psi 
_{i}^{\left( + \right)} \left( {{\rm {\bf r}},t} \right)$ and $\psi 
_{i}^{\left( - \right)} \left( {{\rm {\bf r}},t} \right)$ with asymptotics
\begin{equation}
	\label{eq12}
	\psi_{i}^{\left( + \right)} \left( {{\rm {\bf r}},t_{in} } \right)=\varphi 
	_{i}^{in} \left( {{\rm {\bf r}}} \right)e^{-i\varepsilon_{i}^{in} t_{in} 
	},\,\,\,\psi_{i}^{\left( - \right)} \left( {{\rm {\bf r}},t_{out} } 
	\right)=\varphi_{i}^{out} \left( {{\rm {\bf r}}} \right)e^{-i\varepsilon 
		_{i}^{out} t_{out} }.
\end{equation}
Here, $\varphi_{i}^{in} \left( {{\rm {\bf r}}} \right),\,\varphi_{i}^{out} 
\left( {{\rm {\bf r}}} \right)$ are eigenfunctions of the Dirac Hamiltonian at the appropriate time points
\begin{equation}
	\label{eq13}
	H\left( {t_{in} } \right)\varphi_{i}^{in} \left( {{\rm {\bf r}}} 
	\right)=\varepsilon_{i}^{in} \varphi_{i}^{in} \left( {{\rm {\bf r}}} 
	\right),
\end{equation}
\begin{equation}
	\label{eq14}
	H\left( {t_{out} } \right)\varphi_{i}^{out} \left( {{\rm {\bf r}}} 
	\right)=\varepsilon_{i}^{out} \varphi_{i}^{out} \left( {{\rm {\bf r}}} 
	\right).
\end{equation}
For the given $\kappa $, initial states, including bound ones and the states 
of the continuous spectrum, can be obtained by solving the boundary problem 
for Eq. (\ref{eq8}) with the matrix of $H_{\kappa } \left( {t_{in} } \right)\equiv$ {\texttt H}.

For brevity sake, in (\ref{eq10}) and below, the state $i$ refers to the state with 
specified $\kappa ,\,\,m_{\varphi } $ (see Eq. (\ref{eq10})).

For solving Eq. (\ref{eq11}) within the time interval $\left( {t_{in} ,\,t_{out} } 
\right)$, let us carry out the expansion of $\Phi \left( {r,t} \right)$ in 
terms of the basis of eigenstates of a matrix {\texttt H}, taking into account 
initial condition (\ref{eq11}).
\begin{equation}
	\label{eq15}
	\Phi_{i} \left( {r,t} \right)=\sum\limits_{k=1}^N {c_{ki} \left( t 
		\right)u_{k} \left( r \right)} \,\varepsilon^{-i\varepsilon_{k} t}.
\end{equation}
In, (\ref{eq15}) $N$ is the number of states, $\varepsilon_{k} ,\,u_{k} \left( r 
\right)$ are eigenvalues and eigenfunctions of the matrix $H,c_{ki} $ are 
coefficients of expansion.In this case
\begin{equation}
	\label{eq16}
	c_{ki} \left( {t_{in} } \right)=\delta_{ki} .
\end{equation}
Coefficients $c_{ki} \left( t \right)$ are obtained from the equation:
\begin{equation}
	\label{eq17}
	i\frac{\partial }{\partial t}c_{ji} \left( t \right)=\sum\limits_k {V_{jk} } 
	\left( t \right)c_{ki} \left( t \right),
\end{equation}
where
\begin{equation}
	\label{eq18}
	V_{jk} \left( t \right)=\left\langle {u_{j} \left| {\left( {V^{\left( {mon} 
					\right)}\left( {r,t} \right)-V^{\left( {mon} \right)}\left( {r,t_{in} } 
				\right)} \right)} \right|u_{k} } \right\rangle e^{-i\left( {\varepsilon_{k} 
			-\varepsilon_{j} } \right)t}.
\end{equation}
As a result, in Refs. \cite{bib35} - \cite{bib40}, the solutions of the Dirac equation $\psi 
^{\left( + \right)}\left( {{\rm {\bf r}},t} \right),\,\,\psi^{\left( - 
	\right)}\left( {{\rm {\bf r}},t} \right)$ are numerically determined with 
variations in $t$ within the interval of $\left[ {t_{in} ,t_{out} } \right]$.

In Refs. \cite{bib35} - \cite{bib40}, the 
authors use the formalism ''in''and ''out'' of the vacuum states with the 
introduction of the appropriate annihilation operators for the particles
\begin{equation}
	\label{eq19}
	\left.\left. \left.  {\hat{{b}}_{i}^{in} } \right|0,in \right\rangle \,\,\,=0,\,\,\,\left. 
	{\hat{{b}}_{i}^{out} } \right|0,out \right\rangle \,\,\,=0,\,\,\,i>F.
\end{equation}
Similarly for the anti-particles
\begin{equation}
	\label{eq20}
	\left.\left. \left.  {\hat{{d}}_{i}^{in} } \right|0,in \right\rangle \,\,\,=0,\,\,\,\left. 
	{\hat{{d}}_{i}^{out} } \right|0,out \right\rangle \,\,=0,\,\,\,i<F.
\end{equation}
In (\ref{eq19}), (\ref{eq20}), the capital letter $F$ denotes the Fermi level with the energy
\begin{equation}
	\label{eq21}
	\varepsilon_{F} =-m.
\end{equation}
The electron-positron field operator $\hat{{\psi }}\left( {{\rm {\bf r}},t} 
\right)$ in the Heisenberg's representationis determined as
\begin{equation}
	\label{eq22}
	\hat{{\psi }}\left( {{\rm {\bf r}},t} \right)=\sum\limits_{i>F} 
	{\hat{{b}}_{i}^{in} } \psi_{i}^{\left( + \right)} \left( {{\rm {\bf r}},t} 
	\right)+\sum\limits_{i<F} {\hat{{d}}_{i}^{in\,\,\dag } } \psi_{i}^{\left( + 
		\right)} \left( {{\rm {\bf r}},t} \right),
\end{equation}
\begin{equation}
	\label{eq23}
	\hat{{\psi }}\left( {{\rm {\bf r}},t} \right)=\sum\limits_{i>F} 
	{\hat{{b}}_{i}^{out} } \psi_{i}^{\left( - \right)} \left( {{\rm {\bf r}},t} 
	\right)+\sum\limits_{i<F} {\hat{{d}}_{i}^{out\,\,\dag } } \psi_{i}^{\left( 
		- \right)} \left( {{\rm {\bf r}},t} \right).
\end{equation}
The number of electrons $n_{k} $, created in states $k>F$, is determined by the 
following expressions:
\begin{equation}
	\label{eq24}
	n_{k} =\left\langle {0,\,\,in\left| {\hat{{b}}_{k}^{out\,\,\dag } 
			\hat{{b}}_{k}^{out} } \right|0,\,\,in} \right\rangle =\sum\limits_{i<F} 
	{\left| {a_{ik} } \right|^{2},} 
\end{equation}
where
\begin{equation}
	\label{eq25}
	a_{ik} \left( t \right)=\int {d{\rm {\bf r}}\psi_{i}^{\left( - \right)\dag 
		} \left( {{\rm {\bf r}},t} \right)\,} \psi_{k}^{\left( + \right)} \left( 
	{{\rm {\bf r}},t} \right)=a_{ik} .
\end{equation}
Amplitudes $a_{ik}$ in (\ref{eq25}) do not depend on time and can be written as
\begin{equation}
	\label{eq26}
	a_{ik} =\int {d{\rm {\bf r}}\varphi_{i}^{out\,\,\dag } \left( {{\rm {\bf 
					r}}} \right)e^{i\varepsilon_{i}^{out} t_{out} }\,} \psi_{k}^{\left( + 
		\right)} \left( {{\rm {\bf r}},t_{out} } \right)=\int {d{\rm {\bf r}}\psi 
		_{i}^{\left( - \right)\dag } \left( {{\rm {\bf r}},t_{in} } \right)\,\varphi 
	}_{k}^{in} \left( {{\rm {\bf r}}} \right)e^{-i\varepsilon_{i}^{in} t_{in} 
	}.
\end{equation}
The total number of electron-positron pairs is
\begin{equation}
	\label{eq27}
	P_{st} =\sum\limits_{k>F} {n_{k} } .
\end{equation}

In compliance with the above formalism, the creation probabilities of pairs 
as a function of $Z_{\Sigma } $ were calculated for the central collisions 
of $R_{\min } =7.5fm$ (see Table 1).

\begin{table}[htbp]
	\begin{center}
	\begin{tabular}{|c|c|c|c|c|}
		\hline
		$Z_{\Sigma } $ & $140$ & $150$ & $160$ & $170$ \\
		\hline
		$P_{st} $& $2.16 \cdot 10 ^{-6}$ & $1.69 \cdot 10^{-5}$ & $1.31 \cdot 10^{-4}$ &  $9.65 \cdot 10^{-4}$ \\
		\hline
		$P_{st} -\Delta_{1} $& -  & $3.86 \cdot 10^{-6}$ & $2.13 \cdot 10^{-5}$ & $1.59 \cdot 10^{-4}$ \\ 
		\hline
		\end{tabular}
		\end{center}
	\caption{The calculated values of the probability to create electron-positron pairs.}
\end{table}

In Table 1, the probabilities of pair creation are also presented without 
accounting for the contribution of negative-energy states $1S_{1/2}$, 
(see Sec. 5).

The curve $P_{st} \left( {Z_{\Sigma } } \right)$ within the interval of 
$Z_{\Sigma } =140\div 171$ is given in Fig. 4.

The dependence $P_{st} \left( {Z_{\Sigma } } \right)$ is a monotonically 
increasing curve, changing by approximately a factor of 500 over the 
considered range of values $Z_{\Sigma } $.


\begin{figure}[h!]
	\centerline{\includegraphics[width=0.6\linewidth]{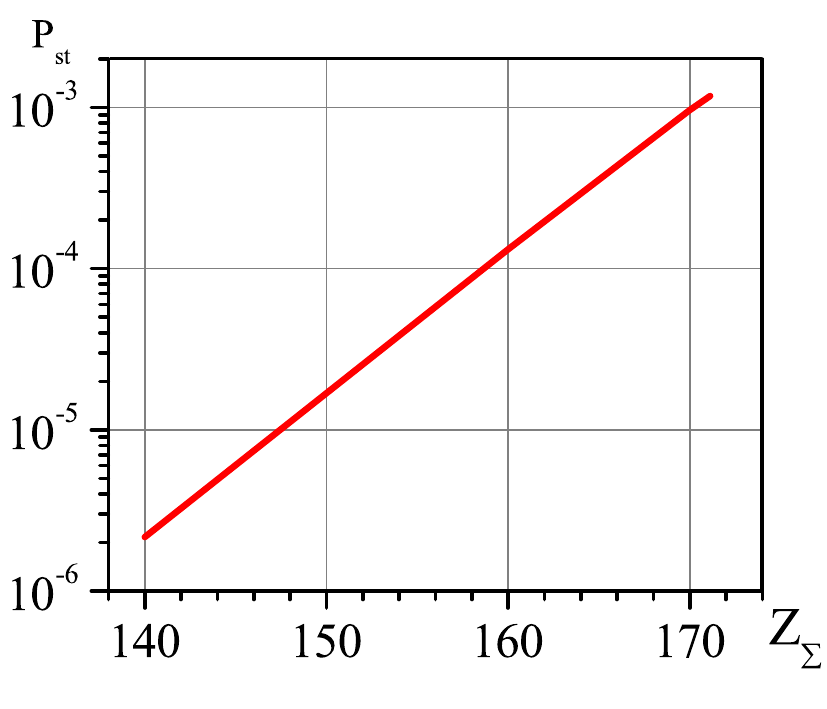}}
	\caption{The creation probability of pairs as a function $Z_{\Sigma } $.}
	\label{ris:Fig.4}
\end{figure}


\section{Creation of electron-positron pairs in the nonperturbative QED in 	the FG and FW representations}
\label{sec:mylabel2}
Within the framework of the perturbation theory in QED in the 
FW representation, when calculating physical effects in 
S-matrix elements, it is suffice to use states with positive fermionic 
energies. It refers to both real and virtual states. The states with 
negative energies are taken into account only in expansions of field 
operators and wave functions in order to meet the conditions of completeness 
and orthonormality. For electrons and positrons with positive energies, 
separate equations are used.

In case of the nonperturbative QED in the FG and 
FW representations, we will proceed from the same premises. 
For electrons, we should use the Eq. (\ref{eq1}), for positrons -- the Eq. 
(\ref{eq3}). The continuous energy spectra of Eqs. (\ref{eq1}) and (\ref{eq3}) as function of 
$Z_{\Sigma } $ are presented in Fig. 3. The discrete energy spectrum of the 
low-lying levels of Eq. (\ref{eq1}) for electrons with positive energies are 
presented ibidem.

In the spectra of Fig. 3, discrete levels with negative energies are not 
available.

At present, the formalism for calculating the creation probability of 
electron-positron pairs, based on the solutions of Eqs. (\ref{eq1}) and (\ref{eq3}) with 
the energy spectra of Fig. 3, has not been developed as yet. However, 
taking into account the equality of Eqs. (\ref{eq2}) and (\ref{eq3}), we can use the 
equivalent continuous electronic spectrum with negative energies determined 
by Eq. (\ref{eq2}) instead of the continuous positron spectrum of Eq. (\ref{eq3}).

In this case, for the QED versions in the FG and 
FW representations, to calculate the probability of pair 
creation, we can use the methodology of the previous section up to the 
formulas (\ref{eq24}) - (\ref{eq27}).

For standard QED, the number of pairs produced in specific states $k>F$ is 
determined by time-independent transition amplitudes from the 
negative-energy continuum of the Dirac equation. 
\[
n_{k} =\sum\limits_{i<F} {\left| {a_{ik} } \right|^{2}.} 
\]
In the standard QED, the total number of electron-positron pairs is
\begin{equation}
	\label{eq28}
	P_{st} =\sum\limits_{k>F} {n_{k} .} 
\end{equation}
In the (QED)$_{FG}$ and (QED)$_{FW}$ versions, transitions labeled by arrows 
d and e in Fig. 2 are absent, so we must exclude the contribution of 
discrete states with negative energies from the summation. As a result, the 
total number of electron-positron pairs is reduced
\begin{equation}
	\label{eq29}
	P_{FG} =P_{FW} =\sum\limits_{k>F} {n_{k} -\Delta_{1} -\Delta_{2} .} 
\end{equation}
Here, $\Delta_{1} $ is the contribution of negative --energy states $1S_{1/2}$ 
at $171>Z>146$; $\Delta_{2} $ is the contribution of negative-energy states 
$2P_{1/2}$ at $184>Z>168$.


\begin{figure}[h!]
	\centerline{\includegraphics[width=0.6\linewidth]{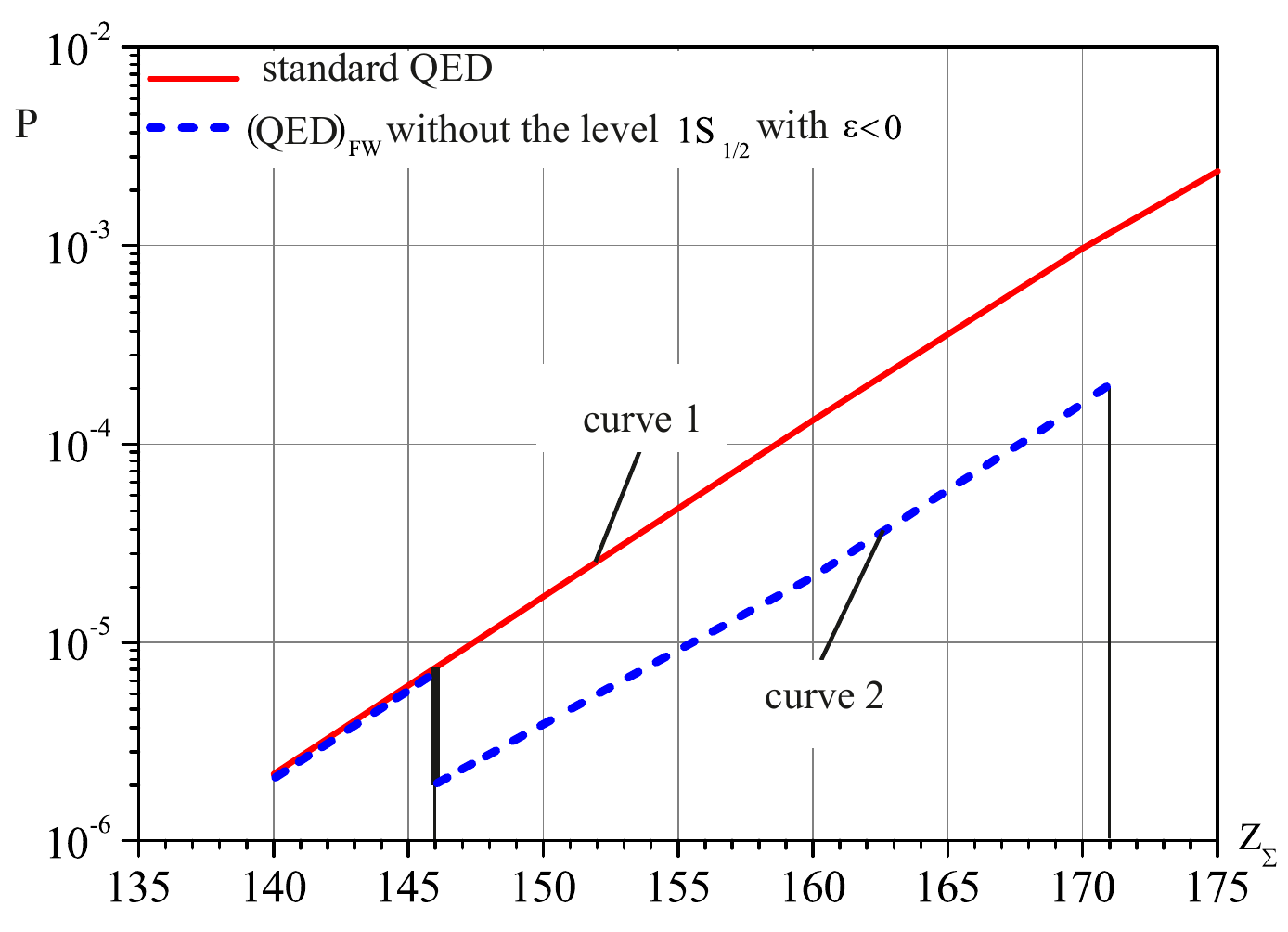}}
	\caption{The creation probability of pairs as a function $Z_{\Sigma } $.}
	\label{ris:Fig.5}
\end{figure}


In Table 1 and in Fig.5 for the interval $Z_{\Sigma } =140\div 170$, the 
creation probabilities of electron-positron pairs $P\left( {Z_{\Sigma } } 
\right)$ are presented both with taking into account and without the 
contribution of $\Delta_{1} $. One can see that the relationship $P_{FW} 
\left( {Z_{\Sigma } } \right)$ has the step nature and differs by a factor of 
several times from the relationship $P_{st} \left( {Z_{\Sigma } } \right)$. 
Evidently, such a difference can be recorded in experiments involving 
collisions of ionized nuclei with precision measurements of the number of 
created electron-positron pairs. 

The data in Table 1 and the dependencies in Figs. 4 and 5 were obtained for 
central collisions with $R_{\min } =17.5fm$. Similar dependencies can be 
obtained for nonzero values of the impact parameter.

When comparing the experimental dependency $P_{\exp } \left( {Z_{\Sigma } } 
\right)$ with the calculated dependencies in Fig. 5, it is necessary to add 
to the latter the calculated dependency for pair production $P_{C} \left( 
{Z_{\Sigma } } \right)$ due to Coulomb excitation of heavy nuclei (see, for 
example, Ref. \cite{bib41} and \cite{bib42})\footnote{My attention was drawn to this 
	significant source of pair production by S.~Yu.~Sedov.}. Dependency $P_{C} 
\left( {Z_{\Sigma } } \right)$ is the same for all considered versions of 
QED.

Table 1 and Figs. 4 and 5 show the probabilities of pair production 
calculated with and without the contribution of negative-energy state $1S_{1/2}$. 
The dependence is step-like and differs significantly from the smooth 
dependence of standard QED. This difference can be detected in experiments 
with precise measurements of the number of produced electron-positron pairs.

\section{Conclusions}
In the paper, it is proposed to carry out a set of experiments with the 
collisions of heavy ions with the total $Z_{\Sigma } =Z_{1} +Z_{2} $ within 
the interval of $146\div 184$. The aim of the experiments to measure the 
number of emitted electron-positron pairs $\left( {P\left( {Z_{\Sigma } } 
	\right)} \right)$ as a function of $Z_{\Sigma } $.

For the standard quantum electrodynamics (QED) with nonempty fermionic 
vacuum, the relationship $P\left( {Z_{\Sigma } } \right)$ is a smooth 
continuous function. For the QED versions with empty fermionic vacuum, 
thenumbers $P\left( {Z_{\Sigma } } \right)$ become smaller. The dependence 
$P\left( {Z_{\Sigma } } \right)$ is a step function. The reasons for such a 
behavior of the relationships $P\left( {Z_{\Sigma } } \right)$ are the 
differences in the contributions of the negative-energy levels.

For the standard QED within the interval of $Z_{\Sigma } =147\div 170$, 
there exists a negative- energy level $1S_{1/2}$ and in the interval of $Z_{\Sigma 
} =169\div 183$, there exists a negative-energy level $2P_{1/2}$. For the QED 
versions with empty fermionic vacuum (the (QED)$_{FG}$, (QED)$_{FW}$ are 
among them), we deal only with positive energies and there is no 
contribution of the above levels $1S_{1/2}$ and $2P_{1/2}$ . It leads to the 
essential decrease in the number of emitted electron-positron pairs within 
the interval $Z_{\Sigma } =147\div 183$ and to the stepwise nature of the 
relationship $P\left( {Z_{\Sigma } } \right)$.

Evidently, such a difference can be recorded in the experiments involving 
collisions of ionized nuclei with precision measurements of the number of 
created electron-positron pairs.

The absence of physical contribution of negative-energy levels in the 
(QED)$_{FG}$ and (QED)$_{FW}$ versions is strictly shown by the analysis of 
Eqs. (\ref{eq1}) - (\ref{eq3}).

The results of the proposed experiments are inherently fundamental and 
essential for the further development of the quantum theory of the field.

In the case of the experimental validation of curve 2 in Fig. 5, the QED 
vacuum is empty; in the QED, there are no processes of vacuum creation of 
pairs; in the QED, there is no Schwinger effect, etc.

In case of the experimental validation of curve 1 in Fig. 5, the vacuum of 
the standard QED is nonempty; in the QED, processes of vacuum creation of 
pairs and the Schwinger effect in a strong electrical field are available. 
The implementation of curve 1 in the experiments will be a reliable 
foundation for experimental validation of neutral vacuum disintegration at 
$Z_{\Sigma }^{cr} \approx 171$ with emission of two positrons.

\section*{Acknowledgments}

This study was conducted within the framework of the scientific program of 
the National Center for Physics and Mathematics, Section ''Particle Physics 
and Cosmology. Stage 2023-2025''. 

The author is thankful to the team led by V.~M.~Shabayev (Saint-Petersburg 
university) for multiple discussions, the calculations to determine the 
relationships for Figs. 4 and 5, the constructive proposals for the future 
research.

The author thanks A.~L.~Novoselova for the essential technical assistance in 
the preparation of the paper.

\section*{Appendix A. FG and FW representations}

Let us write the Dirac equation in an external electromagnetic field in 
covariant form

$$	
	\left( {\gamma^{0}\left( {p^{0}-eA^{0}} \right)-{\rm {\bm \gamma \bm \pi }}-m} 
	\right)\psi_{D} ({\rm {\bf x}},t)=0.\eqno(\rm {A.1})
$$

Multiply (A.1) on the left by an operator with the reversed sign of the 
fermion mass. 
$$
	\left( {\gamma^{0}\left( {p^{0}-eA^{0}} \right)-{\rm {\bm \gamma \bm \pi }}+m} 
	\right)\left( {\gamma^{0}\left( {p^{0}-eA^{0}} \right)-{\rm {\bm \gamma \bm \pi 
		}}-m} \right)\psi_{D} ({\rm {\bf x}},t)=0.
	 \eqno(\rm {A.2})
$$
As a result, we obtain a second-order equation with solutions degenerate 
with respect to the sign of the mass $m$

$$
	\left[ {\left( {p^{0}-eA^{0}} \right)^{2}-\left( {{\rm {\bf p}}-e{\rm {\bf 
					A}}} \right)^{2}-m^{2}+e{\rm {\bf \Sigma H}}-ie{\bm \alpha} {\rm {\bf E}}} 
	\right]\psi_{D} ({\rm {\bf x}},t)=0. \eqno(\rm {A.3})
	$$
Here, ${\rm {\bf H}}=rot{\rm {\bf A}},\,\,{\rm {\bf E}}=-\frac{\partial {\rm 
		{\bf A}}}{\partial t}-\nabla A^{0}$ are the magnetic and electric fields.

Below, we will consider the case of static electromagnetic fields, when 
$p^{0}\psi_{D} =\varepsilon \psi_{D} $.

To transition to the FG representation, it is necessary to use Dirac 
matrices in the chiral representation in Eq. (A.3). This is achieved 
through a unitary transformation

$$
S=\,S^{-1}=\frac{1}{\sqrt 2 }\left( {{\begin{array}{*{20}c}
			I \hfill & \,\,\,\,I \hfill \\
			I \hfill & {-I} \hfill \\
\end{array} }} \right),\eqno(\rm {A.4})
$$
$$
	\psi_{FG} \left( {{\rm {\bf x}},t} \right)=S\psi \left( {{\rm {\bf x}},t} 
	\right)=\left( {{\begin{array}{*{20}c}
				{\varphi_{FG} \left( {{\rm {\bf x}}} \right)} \hfill \\
				{\chi_{FG} \left( {{\rm {\bf x}}} \right)} \hfill \\
	\end{array} }} \right)e^{-i\varepsilon t},\eqno(\rm {A.5})
$$
$$
	\alpha_{c}^{i} =S\alpha^{i} S^{-1}=\left( {{\begin{array}{*{20}c}
				{\sigma^{i}} \hfill & \,\,\,\,0 \hfill \\
				0 \hfill & {-\sigma^{i}} \hfill \\
	\end{array} }} \right),\;\,\,\Sigma_{c}^{i} =S\Sigma^{i} S^{-1}=\left( 
	{\begin{array}{l}
			\sigma^{i}\;\;0\,{\kern 1pt} \\ 
			0\;\;\,\,\sigma^{i} \\ 
	\end{array}} \right).\eqno(\rm {A.6})
$$
In the FG representation, Eq.(A.3) does not mix the upper and lower 
components of the bispinor $\psi_{FG} \left( {{\rm {\bf x}},t} \right)$. In 
the case of stationary states, Eq. (A.3) reduces to two separate 
equations for the spinors $\varphi_{FG} \left( {{\rm {\bf x}}} \right)$ and 
$\chi_{FG} \left( {{\rm {\bf x}}} \right)$.

$$
	\left[ {\left( {\varepsilon -eA^{0}} \right)^{2}-\left( {{\rm {\bf p}}-e{\rm 
				{\bf A}}} \right)^{2}-m^{2}+e{\rm {\bm \sigma {\bf H}}}-ie{\rm {\bm \sigma {\bf E}}}} 
	\right]\varphi_{FG} ({\rm {\bf x}})=0,\eqno(\rm {A.7})
$$
$$
	\left[ {\left( {\varepsilon -eA^{0}} \right)^{2}-\left( {{\rm {\bf p}}-e{\rm 
				{\bf A}}} \right)^{2}-m^{2}+e{\rm {\bm \sigma {\bf H}}}+ie{\rm {\bm \sigma {\bf E}}}} 
	\right]\chi_{FG} ({\rm {\bf x}})=0.\eqno(\rm {A.8})
$$

Equations (A.7) and (A.8) represent one of the variants of fermionic 
Klein-Gordon-type equations with spinor wave functions.

Similar equations were previously considered by Feynman and 
Gell-Mann \cite{bib43}.

It is noteworthy that Eqs. (A.7) and (A.8) are related to the equations 
in the FW representation \cite{bib34}. This 
connection automatically resolves the problem of ''extraneous'' solutions. 
Eq. (A.7) should be used for positive energies $\varepsilon =\left| E 
\right|>0$. In this case, $\varphi_{FG} \left( {{\rm {\bf x}}} 
\right)=A_{\left( + \right)} \,\varphi_{c} \left( {{\rm {\bf x}}} \right)$, 
where $\varphi_{c} \left( {{\rm {\bf x}}} \right)$ is the upper spinor in 
the FW representation $\left( {\psi_{FW}^{\left( + \right)} 
	({\rm {\bf x}})=\left( {{\begin{array}{*{20}c}
				{\varphi_{c} ({\rm {\bf x}})} \hfill \\
				\,\,\,\,\,\,0 \hfill \\
	\end{array} }} \right)} \right)$. Equation (A.8) should be used for negative 
energies $\varepsilon =-\left| E \right|<0$. In this case, $\chi_{FG} 
\left( {{\rm {\bf x}}} \right)=A_{\left( + \right)} \,\chi_{c} \left( {{\rm 
		{\bf x}}} \right)$, where $\chi_{c} \left( {{\rm {\bf x}}} \right)$ is the 
lower spinor in the FW representation $\left( {\psi_{FW}^{\left( - \right)} 
	({\rm {\bf x}})=\left( {{\begin{array}{*{20}c}
				\,\,\,\,\,\,0 \hfill \\
				{\chi_{c} ({\rm {\bf x}})} \hfill \\
	\end{array} }} \right)} \right)$.


\end{document}